\documentclass[onecolumn]{article}
\usepackage{graphicx}
\usepackage{amsmath}

\input{tcilatex}

\begin{document}

\title{Crash Avoidance in a Complex System}
\author{Michael L. Hart$^{\text{*}}$, David Lamper and Neil F. Johnson \\
%EndAName
Physics Department, Oxford University, Oxford OX1 3PU, U.K.}
\maketitle

\begin{abstract}
Complex systems can exhibit unexpected large changes, e.g. a crash in a
financial market. We examine the large endogenous changes arising within a
non-trivial generalization of the Minority Game: the Grand Canonical
Minority Game (GCMG). Using a Markov Chain description, we study the many
possible paths the system may take. This `many-worlds' view not only allows
us to predict the start and end of a crash in this system, but also to
investigate how such a crash may be avoided. We find that the system can be
`immunized' against large changes: by inducing small changes today, much
larger changes in the future can be prevented.

\vskip\baselineskip
\vskip\baselineskip
\vskip\baselineskip

\noindent $PACS$: 01.75.+m; 02.50.Le; 05.40.+j; 87.23.Ge

\noindent $Keywords$: Complex adaptive systems; Econophysics; Agent-based
models

\noindent $^{\text{*}}$ Corresponding author. Tel +44-1865-272356. Fax
+44-1865-272400.

\noindent E-mail address: michael.hart@physics.ox.ac.uk (Michael L. Hart)
\end{abstract}

\newpage

\section{Introduction}

The ability to produce large, unexpected changes represents one of the most
remarkable dynamical features of complex systems \cite{johansen01}. Examples
include crashes in financial markets, jams in traffic and data networks, and
the collapse of the immune system in humans. The large changes arising in
real-world systems might be exogenous and hence triggered by some event in
the environment which is external to the system - however the most
intriguing case concerns the \emph{endogenous} large changes which seem to
arise `out of nowhere' and cannot be blamed on any particular external
cause. The bursting of the dot-com bubble around April 2000 represents a
recent finance-related example of an endogenous large change. The practical
motivation for studying large changes or so-called `extreme events' is
obvious: they are rare, but the damage they cause can be catastrophic.
Present practice in assessing the risk of such events is purely statistical 
\cite{EVT}. This has two shortcomings. First, these events are rare and
hence reliable statistics are difficult to determine. Second, large events
such as crashes arise from subtle temporal correlations in the time-series
and hence can last over many time-steps. Such changes are therefore not
easily captured by examining probability distribution functions involving a
fixed number of time-steps ~\cite{johansen01}. The academic motivation is
equally strong, since large changes provide a visible demonstration of the
system's internal dynamics and correlations. Reference~\cite{johansen01}
quotes Bacon from Novum Organum: ``Whoever knows the ways of Nature will
more easily notice her deviations; and, on the other hand, whoever knows her
deviations will more accurately describe her ways''. In short, the largest
changes will tend to `scrape the barrel' in some way: hence they may provide
new insight into the structure of the barrel (i.e. the system) and contents
(i.e. the constituent parts and their interactions).

In this paper, we examine large changes using a simple, yet highly
non-trivial, model for a complex system comprising a population of objects
or `agents' repeatedly competing for a limited global resource. In
particular, we study the many possible paths the multi-agent system may take
in the vicinity of a possible large change. This `many-worlds' view allows
us not only to predict the start and end of a large change in this system,
but also to investigate how such a large change might be avoided. We find
that the system can be `immunized': by provoking a small response today, the
future robustness of the system against large changes can be boosted.

Agent-based models are becoming a standard tool for investigating complex
system behavior across many disciplines\cite{econophysics}. Typically each
agent has access to a limited set of recent global outcomes of the system;
she then combines this information with her limited strategy set, chosen
randomly at the start of the game (i.e. quenched disorder) in order to
decide an action at the given timestep. These decisions by the $N$ agents
feed back to generate the system's output. The Minority Game (MG) introduced
by Challet and Zhang\cite{challet} offers possibly the simplest paradigm for
such a complex, adaptive system and has therefore been the subject of much
research activity \cite{econophysics}. Most studies of the MG have focused
on a calculation of both time and configuration (i.e. quenched disorder)
averaged quantities - in particular the `volatility' $\sigma $, where $%
\sigma $ is the standard deviation of the number of agents taking a given
decision. Our previous work has shown that the variation of this averaged $%
\sigma $ as a function of memory size $m$ can be quantitatively explained in
terms of `crowd-anticrowd' collective behavior \cite{us,crowd}. In common
with other MG theories \cite{econophysics}, our crowd-anticrowd theory
implicitly assumes both time-averaging and configuration-averaging.

The approach of studying realization-averaged quantities is standard in 
\newline
physics. It yields meaningful results for a wide range of physical systems
because of the natural self-averaging which occurs in many-particle systems
over the time-scales and length-scales of a typical experiment. However it
is arguably of limited use in the study of real-world complex systems. This
is because the observed evolution of a complex system only provides a \emph{%
single} realization of the many possible paths that the dynamics can take. A
financial market, for example, evolves with the knowledge that a crash
happened at a specific moment in the past: there is a definite pre- and
post-crash dynamics which will tend to condition the possible future paths.
For example, the future behavior of the financial markets is dictated in
part by the spectre of the recent dot-com boom and bust. In the context of
human evolution, social revolutions dominate the subsequent dynamics within
a society. Consequently one would gain little insight into the present state
of the world by including all possible scenarios from the past, such as ones
where the dinosaurs did not become extinct or ones where US did not gain
independence. With this in mind, we have recently introduced two alternative
theoretical approaches to analyse the MG, both of which focus on the
dynamics for a \emph{single} realization of the quenched disorder. The first
approach \cite{DMG} views the MG as a stochastically perturbed deterministic
system. Averaging over the sources of stochasticity, in particular the
coin-tosses used to break ties in strategy scores, yields a set of
deterministic mapping equations \cite{DMG}. Hence, in this system any
particular realization of the quenched disorder will give rise to a single
path for the future dynamics. The second approach takes a complementary,
stochastic viewpoint based on a Markov Chain analysis \cite{THMG}. It is
this latter theoretical approach which we will build upon in the present
paper, focusing on the effect that the stochasticity can have in creating
alternative future paths in the system's evolution. A further distinguishing
feature of the present work is therefore its `many worlds' view of the
future evolution of a multi-agent system. The underlying motivation for such
an approach is our desire to understand the degree to which information
about an upcoming large change is already `encoded' in the system, and hence
the degree to which the large change itself might be avoidable.

\vskip\baselineskip

\section{Grand Canonical Minority Game (GCMG)}

The basic MG \cite{econophysics,challet} comprises an odd number of agents $%
N $ (e.g. traders or drivers) choosing repeatedly between option 0 (e.g.
sell or choose route 0) and option 1 (e.g. buy or choose route 1). The
winners are those in the minority group; e.g. sellers win if there is an
excess of buyers, drivers choosing route 0 encounter less traffic if most
other drivers choose route 1. The outcome at each timestep represents the
winning decision, 0 or 1. A common bit-string of the $m$ most recent
outcomes is made available to the agents at each timestep \cite
{econophysics,challet}. The agents randomly pick $s$ strategies at the
beginning of the game, with repetitions allowed. This initial strategy
allocation represents a quenched disorder. Each strategy is a bit-string of
length $2^{m}$ which predicts the next outcome for each of the $2^{m}$
possible histories. After each turn, the agent assigns one (virtual) point
to each strategy which would have predicted the correct outcome. At each
turn of the game, the agent uses the most successful strategy, i.e. the one
with the most virtual points, among her $s$ strategies. Any ties between
equally successful strategies are resolved by the toss of a coin.

Our Grand Canonical Minority Game (GCMG) was first presented in Ref. \cite
{GCMG}. The main generalizations from the basic MG are as follows: (i) an
agent only participates if she possesses a strategy that has a score equal
to, or greater than, a predefined threshold $r$, and (ii) the strategy
performance is only recorded over the previous $\tau $ timesteps. The GCMG
hence incorporates the following real-world effects: (i) agents (e.g.
traders) only participate if they have a certain level of confidence in the
strategies that they possess, and (ii) the agents will tend to forget (or
consider irrelevant) results that lie too far back in history. These
seemingly trivial modifications of the basic MG yield a dynamical system
with surprisingly rich dynamics. These dynamics depend strongly on whether
the participants are fed real (as opposed to random) history strings, and on
the nature of the quenched disorder corresponding to the initial conditions.
We focus on the low $m$ regime because of the richer dynamics, however our
formalism can be applied for all $m$.

The number of agents holding a particular combination of strategies can be
written as a $D\times D\times \dots $ ($s$ terms) dimensional tensor $\Omega 
$, where $D$ is the total number of available strategies. For $s=2$
strategies per agent, this is simply a $D\times D$ matrix where the entry $%
(i,j)$ represents the number of agents who picked strategy $i$ and then $j$.
The strategy labels are given by the decimal representation of the strategy
plus unity, for example the strategy 0101 for $m=2$ has strategy label
5+1=6. This quenched disorder $\Omega $ is fixed at the beginning of the
game and is written using the full strategy space \cite{THMG}. The value of $%
a_{R}^{\mu }$ describes the prediction of strategy $R$ given the history $%
\mu $, where $\mu $ is the decimal number corresponding to the $m$-bit
binary history string \cite{challet}. Hence $a_{R}^{\mu }=-1$ denotes a
prediction of option `0' while $a_{R}^{\mu }=1$ denotes a prediction of
option `1'.

As discussed in Sec. I, we are interested in the detailed dynamics of the
game for a given realization of the initial quenched disorder $\Omega $.
Hence we imagine that a particular $\Omega $ has already been chosen. Since
the game involves a coin-toss to break ties in strategy scores, this
stochasticity also means that different runs for a given $\Omega $ will also
differ - we return to this point below. The number of traders making
decision $1$ (the `instantaneous crowd') minus the number of traders making
decision $0$ (the `instantaneous anticrowd') defines the net `attendance' $%
A[t]$ at a given timestep $t$ of the game. This attendance $A[t]$ is made up
of two groups of traders at any one timestep \cite{THMG}: \vskip%
\baselineskip

\noindent (a) $A_{D}[t]$ traders who act in a `decided' way, i.e. they do
not require the toss of a coin to decide which option to choose. This is
because they have one strategy that is better than their other playable
strategies, or because their highest-scoring strategies are tied \emph{but}
give the same response as each other to the history $\mu _{t}$ at that turn
of the game. \vskip\baselineskip

\noindent (b) $A_{U}[t]$ traders who act in an `undecided' way, i.e. they
require the toss of a coin to decide which option to choose. This is because
they have two (or more) tied, highest-scoring strategies \emph{and} these
give different responses to the history $\mu _{t}$ at that turn of the game. %
\vskip\baselineskip

\noindent Hence the attendance at timestep $t$ is given by 
\begin{equation}
A[t]=A_{D}[t]+A_{U}[t]\ \ \ .
\end{equation}
The state of the game at the beginning of timestep $t$ depends on the
strategy score vector ${\underline{s}}_{t}$\ and history $\mu _{t}$.
Henceforth we will drop the variable $t$ from the notation, but note that it
remains an implicit variable through the time-dependence of $\underline{s}$
and $\mu $. We also focus on $s=2$ strategies per agent, although the
formalism can be generalized in a straightforward way. At timestep $t$, $%
A_{D}$ is given exactly by 
\begin{equation}
A_{D}(\underline{s},\mu )=\sum_{R,R^{^{\prime }}}a_{R}^{\mu }(1+\mathrm{Sgn}%
[s_{R}-s_{R^{^{\prime }}}])\mathrm{H}[s_{R}-r]\underline{\underline{\Psi }}%
_{R,R^{^{\prime }}}
\end{equation}
where the symmetrized matrix $\underline{\underline{\Psi }}=\frac{1}{2}(%
\underline{\underline{\Omega }}+\underline{\underline{\Omega }}^{T})$ with $%
\underline{\underline{\Omega }}$ representing the quenched disorder. The
element $\underline{\underline{\Omega }}_{R,R^{^{\prime }}}$ gives the
number of agents picking strategy $R$ and then $R^{\prime }$. The function $%
\mathrm{H}[\dots ]$ is the Heaviside function. The number of undecided
traders $N_{U}$ is given exactly by 
\begin{equation}
N_{U}(\underline{s},\mu )=\sum_{R,R^{^{\prime }}}\delta
(s_{R}-s_{R^{^{\prime }}})[1-\delta (a_{R}^{\mu }-a_{R^{^{\prime }}}^{\mu })]%
\mathrm{H}[s_{R}-r]\underline{\underline{\Omega }}_{R,R^{^{\prime }}}
\end{equation}
and therefore the attendance of undecided traders $A_{U}$\ is distributed
binomially as follows: 
\begin{equation}
A_{U}(\underline{s},\mu )\equiv 2\ \mathrm{B}[N_{U}(\underline{s},\mu ),%
\frac{1}{2}]-N_{U}(\underline{s},\mu )
\end{equation}
where the term `$\mathrm{B}[N_{U}(\underline{s},\mu ),\frac{1}{2}]$' is
standard notation to denote a binomial distribution in which there are $%
N_{U}(\underline{s},\mu )$ trials with probability $1/2$ of obtaining a
given outcome \cite{binref}. Having obtained $A_{D}(\underline{s},\mu )$\
and $N_{U}(\underline{s},\mu )$, it is now possible to calculate the
probability $P(w=0|\underline{s},\mu )$ that the next winning decision is a
0, given knowledge of $\underline{s}$\ and $\mu $. This probability is given
as follows: 
\begin{equation}
P(w=0|\underline{s},\mu )=(\frac{1}{2})^{1+N_{U}}\sum_{i=0}^{i=N_{U}}\text{ }%
^{N_{U}}C_{i}(1+\mathrm{Sgn}[A_{D}+2i-N_{U}])
\end{equation}
We note that Eqs. (1)-(5) represent a generalization to the GCMG, of the
equations presented in Ref. \cite{THMG} to describe the basic MG within the
Markov Chain formalism.

\vskip\baselineskip

\section{Analysis of a Crash}

The dynamics of the GCMG are rich, and exhibit large changes \cite{GCMG}.
Here we look in detail at a typical (randomly-chosen) large change. In
particular we are going to look at the many paths along which the system can
evolve, starting well before the large change itself.

The quenched disorder is defined by $\Omega $. The exact numerical form of $%
\Omega $ is too cumbersome to show but can be obtained from the authors. The
initial conditions are defined by the score vector $\underline{s}$\ and the
corresponding history state $\mu $. For the purpose of calculating the
values of $\underline{s}$ and $\mu $\ in subsequent timesteps, we also
define the last $\tau +m$\ winning outcomes of the GCMG. The specific
initial conditions employed were taken from a typical run of the game, once
initial transients had disappeared.

Equation (5)\ states the probability that the next winning outcome is a 0,
given knowledge of $\underline{s}$\ and $\mu $. It is possible to go further
than this, by stating the mean attendance given that the attendance is
positive (i.e. winning decision of 0) and/or the mean attendance given that
the attendance is negative (i.e. winning decision of 1). Note that given a\
state of $\underline{s}$\ and $\mu $, $P(w=0|\underline{s},\mu )$ may be
greater than 0 but less than 1, and as such both winning decisions are
possible for this state. If however $P(w=0|\underline{s},\mu )$ has a value
of 0 or 1, then there is only one winning decision via which the complex
system may evolve at that time-step. The system is therefore deterministic
at this time-step.

There are several scenarios for $A_{D}$\ and $N_{U}$\ that can be examined.
If $A_{D}\neq 0$ and $N_{U}=0$\ then the price change is given by the value
of $A_{D}$, hence only one winning outcome is possible. If $A_{D}=0$ and $%
N_{U}=0$, then the winning outcome has equal probability of being 0 or 1. In
this case the recorded attendance is zero: however the two possible winning
outcomes are noted for the purpose of calculating the outcomes for
subsequent timesteps. If $N_{U}\neq 0$ and $N_{U}<|A_{D}|$, then the
attendance $A$ can take $N_{U}+1$ possible values, each with their
associated probabilities calculated using simple binomial theory in a
similar way to Eq. (5). The magnitude of the attendance at $t=\gamma $ has
no significance for the future states of the system at times $t>\gamma $.
Only information concerning the sign of the attendance is carried forward,
via the winning outcome. It is therefore sufficient to record the attendance
for that timestep as being the average possible value, i.e. $A=A_{D}$ if $%
N_{U}\neq 0$ and $N_{U}<|A_{D}|$. Showing all $N_{U}+1$ possible price
changes will not provide us with any extra useful information concerning the
system - instead, showing the average value will enable us to present a less
cluttered image of the many worlds (see later). If $N_{U}\neq 0$ and $%
N_{U}>|A_{D}|$, then both winning outcomes 0 and 1 are possible. For ease of
calculation we shall approximate the binomial distribution in $A_{U}$ using
a Normal distribution. Hence we can extract two possible values for the
attendance: one corresponding to a winning outcome of 0, and the other a
winning outcome of 1. To calculate these attendances, one should split the
Normal distribution into two parts with the dividing line at the point of
cut-off in terms of determining the winning outcome. If the winning outcome
is to be a 0, then the corresponding attendance contributed by $N_{U}$
becomes the first moment of the distribution above the cut-off. Figure 1
shows an example of this scenario, and indicates the two corresponding
attendances that are to be calculated.

We\ define a `path' as the trajectory of the complex system during one
possible run of the game. The game is set up with $N=101$ agents, each with
two $m=3$ strategies scored over the last $\tau =60$ timesteps. The agents
have identical confidence levels given by $r=0.52\ast \tau =31.2$. Figure 2a
shows all 949 different paths in terms of the cumulative attendance (e.g.
price in a financial market). that this complex system can take over 56
timesteps, starting from the given initial conditions. Figure 2b indicates
the probabilities of occurrence of the paths in one run of the game, as a
function of the paths' end points. Figure 2c shows a measure of the spread
in price-increment between all the different paths as a function of time.
This measure of the spread in price-increment is defined as follows: 
\begin{equation}
\lambda =\frac{1}{2}\sum_{i}\left[ \sum_{j}(x_{j}-x_{i})^{2}P(x_{j})\right]
P(x_{i})
\end{equation}
where $x_{\alpha }$ is the price change between neighboring timesteps for a
price path denoted by $\alpha $. The probability that path $x_{\alpha }$\
occurs, given the initial conditions, is given by $P(x_{\alpha })$. We note
that for the basic MG, $\lambda $\ would take a value of $99.4$ if the
attendances were random. In the present GCMG case, the random limit would be
larger due to the fluctuation in the total number of agents playing at a
given time-step.

It becomes very difficult to resolve the different paths in Fig. 2a after
timestep 40, even though it is only the average, historically-distinct paths
which have been shown. The thick line shows the weighted average of \emph{all%
} these paths. [Equation (5) allows us to calculate the probability of each
path being the actual path taken in a single run of the game]. It is easy to
identify by eye that a large change (i.e. so-called drawdown or crash)
occurs between time-steps 33 and 50. Figure 2c shows that $\lambda $\ takes
a very small value at the start of the crash: this signifies that the
price-changes are similar at this point in time, in terms of the direction
and size of movement, for the large majority of separate paths. Similarly at
the end of the crash the quantity $\lambda $ takes a very large value,
signifying a point in time where there is great disparity in the sign and
size of price-changes between separate paths. We note that $\lambda $\ is
zero for the first 12 steps, since there is only one path possible up until
timestep 13. Figure 2c demonstrates that it is possible to identify the
start of a crash ahead of time, by the collective movements of the `parallel
worlds'. In the case shown, the start of the crash is indicated by the
collective crashes of the large majority of the many parallel worlds.
Furthermore, it is possible in this system to identify the point in time
when the system is released from the crash by observing when there is
significant disparity between the parallel-world path movements. Starting
from a time 33 steps prior to the start of the crash, it can be seen that a
crash is highly likely.

Despite being highly likely, such a crash is not however inevitable. In
order to gain insight into how the crash may be managed, reduced or even
avoided, we now turn to investigate the differences between the paths which
perform better/worse than the many-world average. We will take `better than'
as being a path that has an end value that is numerically greater than that
of the mean at timestep 56 (c.f. Fig. 2a) - similarly for `worse than'.
Hence we shall split up the paths into these two corresponding groups:
better-than-the-mean shall be labelled Set A, and worse-than-the-mean shall
be labelled Set B.

Figure 3a shows the probability tree of a path belonging to Set A, for the
first 33 timesteps. Each branching in this figure corresponds to a branching
in Fig. 2a. Measuring the cumulative sign of attendance (i.e. price in a
financial market) at timestep 33, one gets three values for the particular
initial conditions: $0$, $2$ and $4$. It is interesting to segregate those
paths that have a cumulative movement of $0$, $2$, and $4$, and plot the
equivalent graphs to those shown in Fig. 3a together with their weighted
means. Figure 3b shows those paths with a net movement of $0$. Figure 3c
shows those paths with a net movement of $2$. Figure 3d shows those with a
net movement of $4$. The mean probability of ending up as a member of Set A
is \emph{greater} for those paths with a net movement of $2$ as compared to
those with $4$. Similarly, the mean probability of ending up as a member of
Set A is \emph{greater} for those paths with a net movement of $0$ as
compared to those with $2$.

We next compare weighted histograms of the cumulative \emph{sign} of
attendance during a period prior to the crash, and during a period after the
crash. This will reinforce our findings concerning the differences between
the paths in Set A and Set B. To do this, we set the period prior to the
crash as timesteps 11 to 35, and the period after the crash as timesteps 36
to 56. Figure 4a shows the histogram for Set A prior to the crash. Figure 4b
shows the histogram for Set B prior to the crash. It can be seen that the
paths in Set B generally have \emph{greater} cumulative signs of attendance
than the paths in Set A. Figure 4c shows the histogram for Set A after the
crash, while Fig. 4d shows the histogram for Set B also after the crash. Now
the paths in Set B generally have \emph{smaller} cumulative signs of
attendance than the paths in Set A.

We now examine the occurrence of (m+1)-bit words contained in the sequence
of winning outcomes during the period from timestep 11 to 35, for the paths
in Set A and Set B. [Studying the occurrence of (m+1)-bit words is
equivalent to studying the occurrence of a 0 or a 1 following an m-bit
history]. Figure 5 shows the frequency of occurrence of an (m+1)-bit word
for Set A, divided by that for Set B. This indicates how much more likely a
particular (m+1)-bit word is in Set A as compared to Set B, prior to the
crash. Hence it is just over twice as likely that 000 will be followed by a
0 in a path belonging to Set A, as compared to a path belonging to Set B.
The period preceding the crash can therefore be seen as one of `imbalance'
or `tension', which is subsequently released by the crash.

Figure 6 shows an $m=3$ de-Bruijn graph \cite{THMG}. The edges represent
transitions between history states. Transitions having peak values above $%
1.15$ in Fig. 5, are shown with darker arrows. Transitions that are
particularly weak are represented by dashed arrows. The paths which are
members of Set A, i.e. those which perform \emph{better} than the mean, tend
to have more \emph{down} movements in their history prior to the crash. By
contrast, those paths which are members of Set B, i.e. those which perform 
\emph{worse} than the mean, tend to have more \emph{up} movements in their
history prior to the crash. If one therefore wanted to avoid the crash in a
run of the game, then it would be wise to force the system to undergo \emph{%
down} movements prior to the crash, so as to relax the build up of any
unwanted bias in the strategy scores. This would then keep the system to the
left-hand side of the De-Bruijn graph where the arrows are darker, as seen
in Figure 6. A controlling force or `complex systems manager' could
intervene at moments where there is little cost in order to manipulate a
winning outcome, i.e. at time-steps where it may require as little as one
agent to change the path of the system. In this way the controlling force
could avoid the crash with minimal effort or investment.

This result that a large change can be prevented by creating an earlier,
smaller movement \emph{in the same direction}, reminds us of the idea of
`immunization'. Injecting a small amount of a virus provokes a minor
response which `primes' the system's defences against a much larger, and
potentially fatal, response later on. In the same way, a small crash induced
today, can be used to prevent a much larger crash from arising later on.

It has recently been shown that the large endogenous changes which arise in
the GCMG complex system, exhibit a degree of predictability \cite
{DMG,crashDMG}. Information signaling an increased threat of a large change
is `encoded' in the system's microstructure well in advance. The present
results for the GCMG are therefore consistent with these earlier findings 
\cite{DMG,crashDMG}. In real-world complex systems, such encoding would
clearly be far more complicated to unravel. However there is the hope that 
\emph{if} the system is viewed with the right `spectacles', such signals
might be uncovered. We refer the reader to the large body of work of
Sornette et al. (see Ref. \cite{johansen01} and references therein) for the
fascinating proposal of a log-periodic signature prior to large changes in a
range of physical and natural systems, including financial markets.

Finally, we would like to draw attention to the nature of the large change
studied in this paper. This event is described by a convergence of the many
`world lines' which project into the immediate future. This convergence
arises in spite of the stochastic fluctuations which occur along each `world
line'. It is this effect that gives the system its predictability, and
enables us to identify the birth of the crash. We note that if an observer
was limited to knowledge of just a single `world line', then little could be
said about the crash's appearance until it became completely obvious.
Irrespective of the fact that a crash arose, there is arguably a more
fundamental `event' which occurred in the system's dynamics. This `event' is
the seemingly spontaneous convergence of the diversity of world-lines into
the same direction. This convergence underlies the robustness of the
resulting large change and hence the inevitability of the crash - in short
the crash happened because of `fate' as opposed to chance. This leads us to
speculate that an `event' in which a wide range of world-lines converge, may
be fairly universal in complex systems, and Nature in general. In short, the
reality that is observed in the physical world might correspond to robust
`events' involving convergence of world-lines, and as such is relatively
unaffected by external noise/fluctuations.

\vskip\baselineskip

\section{Conclusion}

We have investigated large changes in the Grand Canonical Minority Game
(GCMG). The GCMG provides a simple, yet highly non-trivial, model for a
complex system comprising a population of objects or `agents' repeatedly
competing for a limited global resource. We have focused on the many
possible paths the multi-agent system may take in the vicinity of a possible
large change. This `many-worlds' view allowed us not only to predict the
start and end of a large change, but also to investigate how such a large
change might be avoided. We uncover the remarkable result that the complex
system can be `immunized' in order to protect it against large changes. In
particular we find that by provoking a small response today in the system,
we can protect it against much larger responses in the future.

\pagebreak

\pagebreak

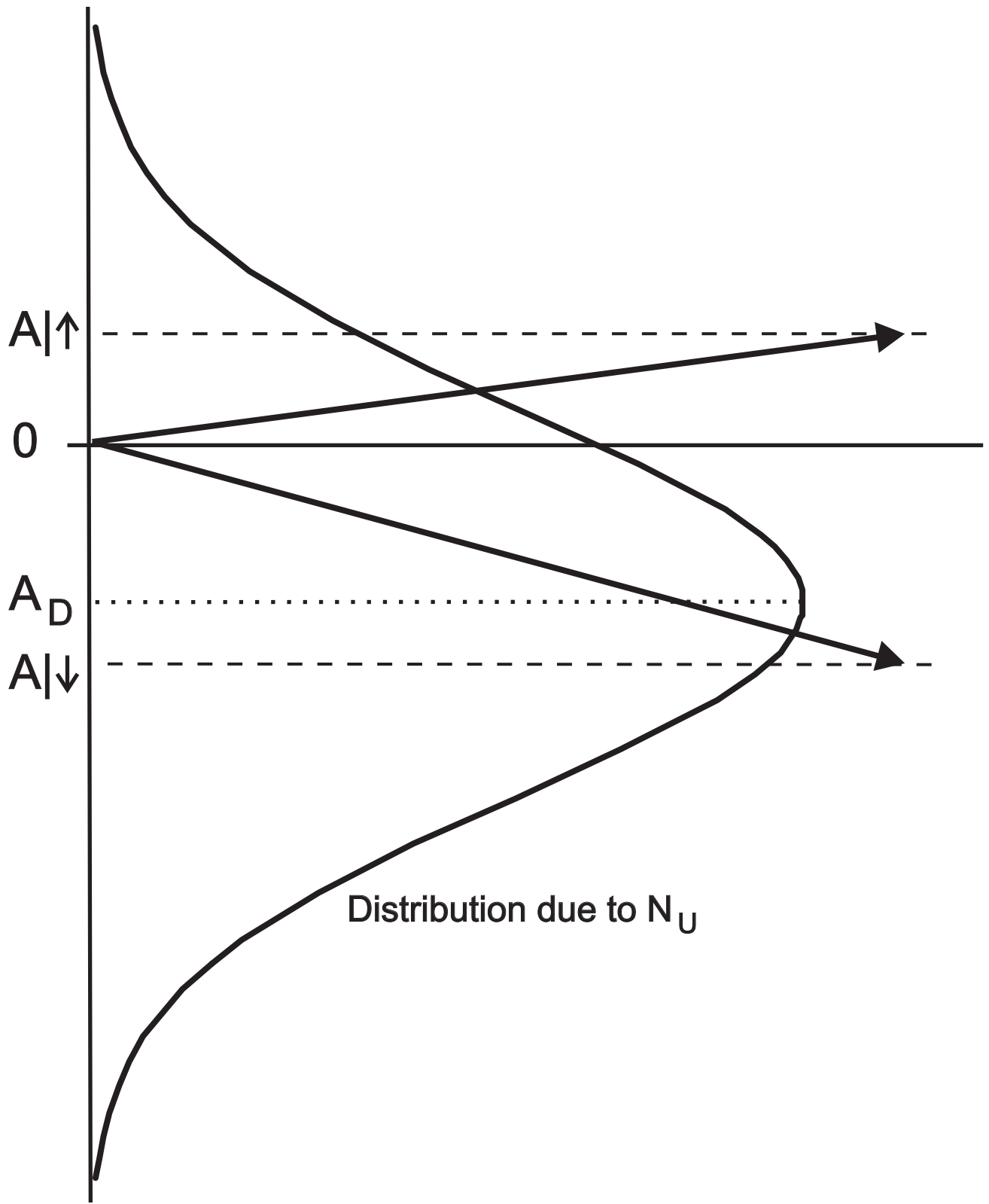
FIG. 1. Demonstration of the two possible values for the attendance given
that $N_{U}\neq 0$ and $N_{U}>|A_{D}|$. Note that the area under the
distribution between $0$ and $A|\uparrow $, is equal to the area between $%
A|\uparrow $\ and the top edge of the distribution. Similarly the area
between $0$ and $A|\downarrow $, is equal to the area between $A|\downarrow $%
\ and the bottom edge of the distribution.

\bigskip

FIG. 2a. The 949 paths that the GCMG\ can take over 56 timesteps around the
crash. The thick line shows the mean path, accounting for the differing
probabilities of the paths.

FIG. 2b. The probabilities for the 949 paths, calculated using Eq. (5).

FIG. 2c. The measure $\lambda $ of the spread in price-increments between
all separate paths, as a function of time. Calculated using Eq. (6).

\bigskip

FIG. 3a. Probability tree for the first 33 timesteps, showing the
probability of a path belonging to Set A, i.e. doing better than the mean at
timestep 56.

FIG. 3b. The paths of the probability tree that have a cumulative movement
of $0$ up to timestep 33. Thick line shows the weighted mean.

FIG. 3c. The paths of the probability tree that have a cumulative movement
of $2$ up to timestep 33. Thick line shows the weighted mean.

FIG. 3d. The paths of the probability tree that have a cumulative movement
of $4$ up to timestep 33. Thick line shows the weighted mean.

\bigskip

FIG. 4a. The weighted histogram of cumulative sign of attendance for Set A,
prior to the crash between timesteps 11 and 35.

FIG. 4b. The weighted histogram of cumulative sign of attendance for Set B,
prior to the crash between timesteps 11 and 35.

FIG. 4c. The weighted histogram of cumulative sign of attendance for Set A,
after the crash between timesteps 35 and 56.

FIG. 4d. The weighted\ histogram of cumulative sign of attendance for Set B,
after the crash between timesteps 35 and 56.

\bigskip

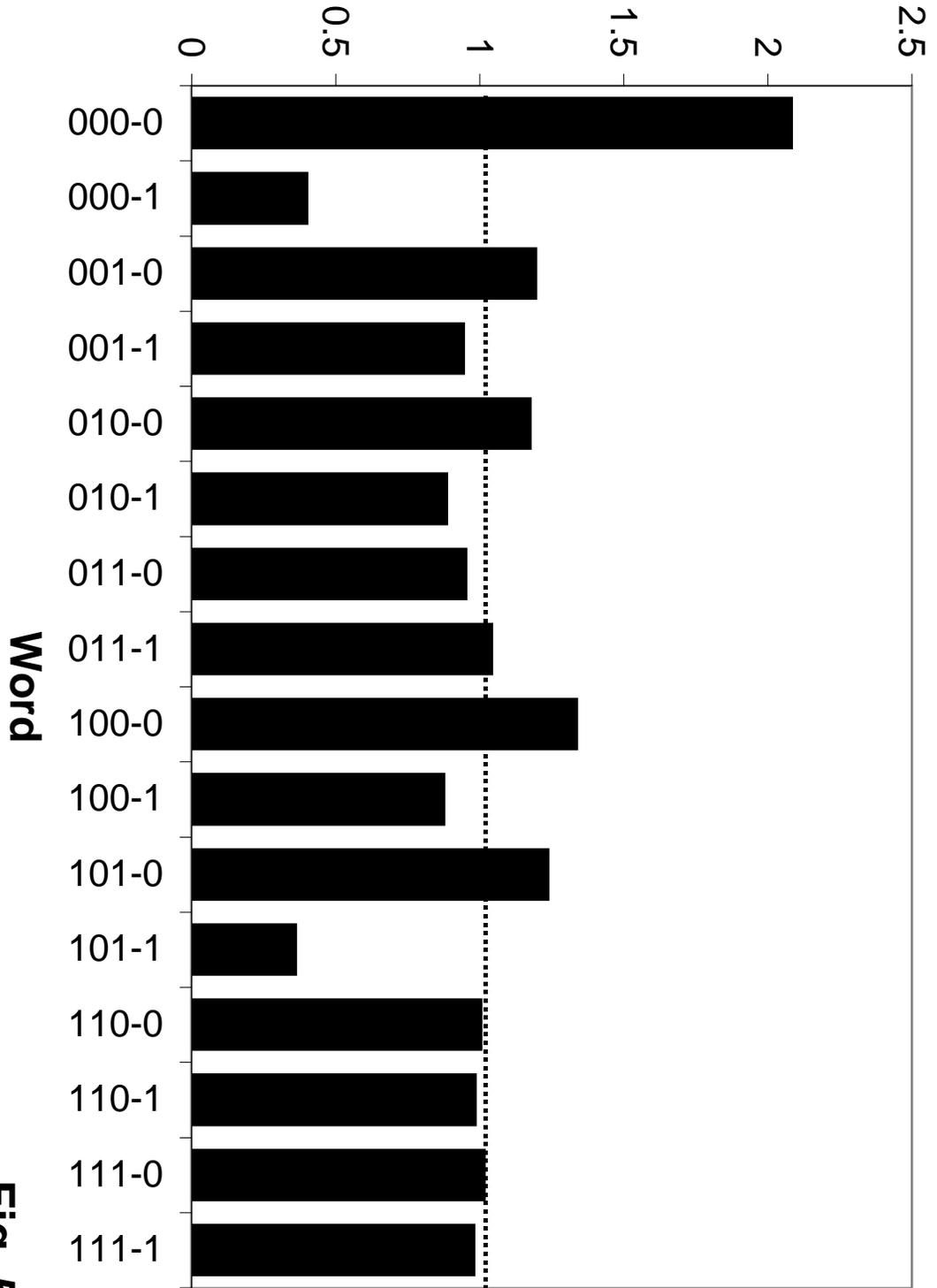
FIG. 5. The (m+1)-bit frequencies for Set A divided by Set B, in the period
prior to the crash between timesteps 11 and 35.

\bigskip

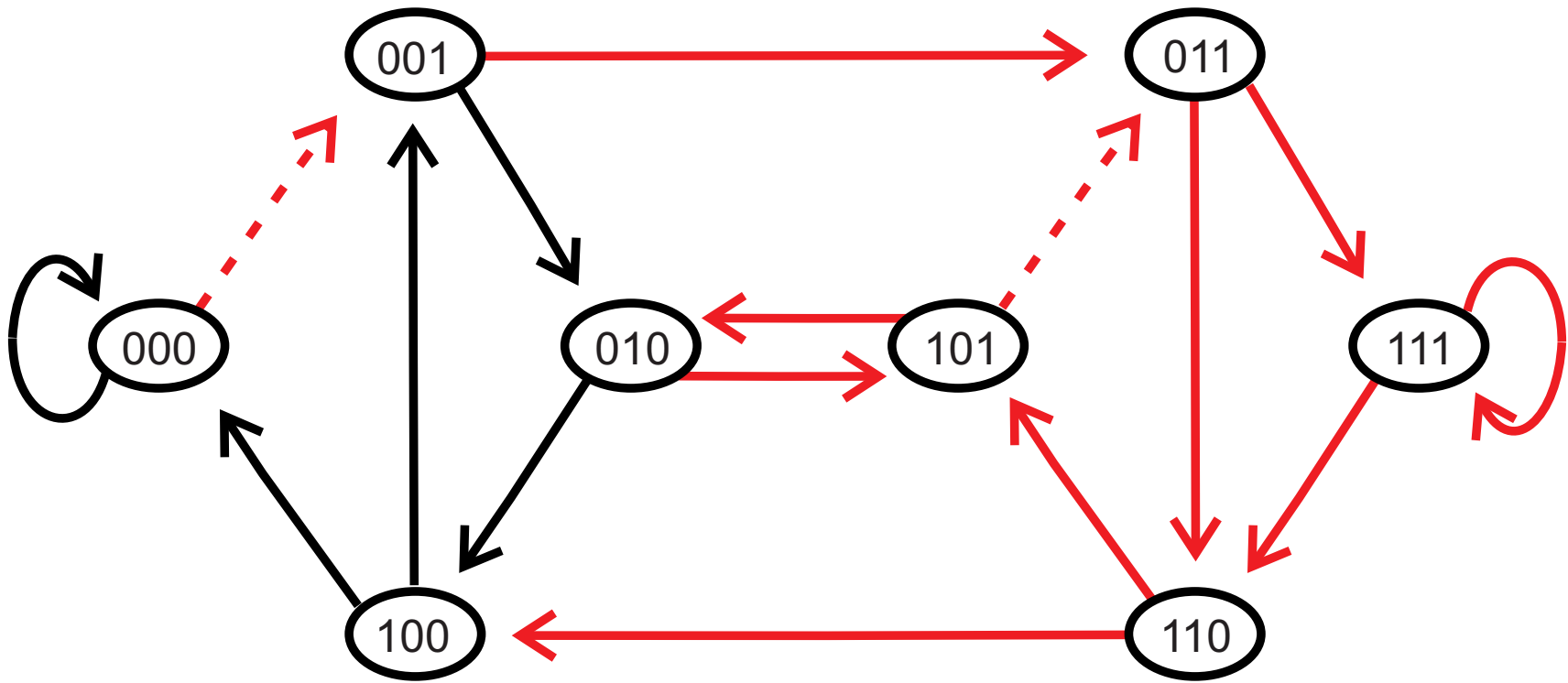
FIG. 6. An $m=3$ De-Bruijn graph. The edges show transitions between history
states. Those edges with peak values above 1.15 in Fig. 5, have darker
arrows. The transitions that are particularly weak are represented by dashed
arrows.

\newpage

\end{document}